\documentstyle[prl,twocolumn,aps]{revtex}
\input{epsf}
\begin{document}
\draft
\twocolumn[\hsize\textwidth\columnwidth\hsize\csname @twocolumnfalse\endcsname
\title{Quantum Computing of Poincar\'e Recurrences and Periodic Orbits}

\author{B. Georgeot}

\address {Laboratoire de Physique Th\'eorique, UMR 5152 du CNRS,
IRSAMC, Universit\'e Paul 
Sabatier, 31062 Toulouse Cedex 4, France}

%\date{\today}
\date{July 30, 2003}

\maketitle

\begin{abstract}
Quantum algorithms are built enabling
to find Poincar\'e recurrence times and periodic orbits of 
classical dynamical systems. 
It is shown that exponential gain compared to classical algorithms can
be reached for a restricted class of systems. Quadratic gain
can be achieved for a larger set of dynamical systems.
The simplest cases can be implemented with small number of qubits.

\end{abstract}
\pacs{PACS numbers: 03.67.Lx, 05.45.Ac, 05.45.Tp}
\vskip1pc]

%\begin{multicols}{2}
\narrowtext

It has been suggested since Feynman \cite{feynman} that the superposition
principle of quantum mechanics enables to perform exponentially many
computations in parallel (see reviews in \cite{josza,steane,nielsen}).
Thus in principle quantum processors using the full power
of quantum mechanics may be enormously faster than classical
nowadays computers. This possibility has motivated a great deal of 
attention in the scientific community, and many experimental
proposals are explored to realize such a quantum computer.
However, it has been surprisingly hard to spot
specific problems where quantum algorithms may be faster than
classical ones.  One such example is the celebrated 
Shor's algorithm \cite{shor} which
factors large numbers with exponential efficiency compare to any known 
classical algorithm. Another algorithm, due to Grover \cite{grover}, 
speeds up the search in an unsorted database,
although not exponentially.  

In parallel, people have investigated the possibility of using a quantum
device to simulate physical systems, a problem
 of much practical
interest.
In particular, several works have shown that quantum computers can speed up
the simulation of quantum mechanical systems \cite{lloyd,BG1,saw1}
and classical
spin systems \cite{lidar}.
It was also suggested in \cite{ascat} that a quantum computer
may be efficient at simulating classical dynamical systems as well. 
In the present work, the problem of finding Poincar\'e recurrence times and
periodic orbits of dynamical systems is studied.
Such quantities
give important information about the dynamics, and are standard tools for
studying classical mechanics.
Explicit algorithms will be presented showing
that such quantities can be evaluated faster on a quantum computer
than on a classical one.  In some cases, 
exponential efficiency compared to classical algorithms
can be reached. In other cases,
quadratic gains can be realized.

For classical bounded conservative systems, Poincar\'e \cite{Poincare}
showed that some points from 
an arbitrary small phase space domain will eventually come back 
to this domain. 
In general, actual values of the recurrence time for
a given area of phase space, and statistics of these times, are 
not given by the theorem and require specific studies to be understood. 
This time can be extremely long, and is very hard to study
numerically.  Advances in the power of modern classical computers
allow to get information about such quantities, but the problem
is still under investigation nowadays \cite{recurrences}. 
In addition to their intrinsic interest, the return times
 give also insight
on transport properties, diffusion coefficients and correlation functions
of the system studied \cite{recurrences}.
A related feature of dynamical systems is the presence of periodic
orbits, which are trajectories coming back exactly to their initial point
in phase space.  They play a special role in understanding
physical properties
of these systems, and are present both in Hamiltonian 
and dissipative systems.
For example, they enable to compute diffusion coefficients
and properties of strange attractors \cite{cvitanovic}, and 
enter the famous Gutzwiller trace formula
\cite{GTF} which is a general semiclassical quantization scheme
valid for chaotic systems.

I will first show how such quantities can be found efficiently
on a particular example, the Arnold cat map 
\cite{arnold,lieberman}, one of the most famous chaotic 
dynamical systems.
It is an automorphism of the torus with equation:

\begin{equation}
\label{catmap}
\bar{y}=y+x \; \mbox{(mod} \;\mbox{1)}\;\;, \;\; \bar{x}=y+2x \;\mbox{(mod} 
\;\mbox{1)}\;,
\end{equation}

where bars denote the new values of the variables after one iteration.
This map is area-preserving, and may be interpreted as describing
the evolution in phase space, with $x$ representing position and 
 $y$ momentum. It is an Anosov system, with homogeneous exponential
divergence of trajectories and positive Kolmogorov-Sinai entropy
$h\approx 0.96$. The map (\ref{catmap}) can be equivalently written 
as action of the $2\times 2$ matrix 
$L=\left( \begin{array} {cc} 2& 1 \\ 1 & 1 \end{array} \right)$
on $\left( \begin{array} {c} x \\ y \end{array} \right)$ .  Other cat maps 
can be built by taking any $2\times 2$ matrix
$L$ with integer entries and determinant $1$.

It is known \cite{hannay,percival,keating} 
that for the cat map (\ref{catmap}) 
periodic orbits are in one-to-one correspondence
with points with rational coordinates.  Still, it is a computationally
hard problem to find the {\em period} of a given point. An usual way
for studying this problem is to consider the set of rational points sharing 
the same denominator $g$. These points form a $g \times g$
lattice in the phase space,
which is invariant under the action of the map.  The action of the map on such
points can be written on the numerators only, namely:
$\bar{y}=y+x \; (\mbox{mod} \;g)\;\;, \;\; \bar{x}=y+2x \;(\mbox{mod} 
\;g)\;,$  or $\left( \begin{array} {c} \bar{x} \\ \bar{y} \end{array} \right)=L\left( \begin{array} {c} x \\ y \end{array} \right) \; (\mbox{mod} \;g)\;\;$, with $x,y,\bar{x},\bar{y}$ integers.

All points in such a lattice are periodic, but not with the same period.
A quantity similar to recurrence time is the
{\em lattice period function } $\alpha (g)$.  It is the smallest integer
such that after $\alpha (g)$ iterations all points in the lattice
have come back to the
initial position, i.e. it is the
recurrence time
of the whole lattice.
 This quantity has
been the subject of many studies \cite{hannay,percival,keating},
since it describes the whole set of periodic orbits of the system, 
and also gives insight on the quantization of the cat map 
\cite{hannay,keating}.  In particular, the behavior of 
$\alpha (g)$ 
controls the ergodicity of the eigenfunctions
of the quantum cat map for $\hbar=1/g$ \cite{rudnick}.
It turns out
to be a very erratic function of $g$;
some results on its statistical properties 
have been derived nonrigorously
based on assumptions from probabilistic number theory \cite{keating}.
Numerical checks for moderate values of $g$ show
that the asymptotic r\'egime is very slowly reached \cite{keating}.
Thus for example computing $\alpha(g)$
in small windows around large values of $g$
will enable to check its asymptotic properties.  This has
interest both for number theory and study of dynamical systems.

A quantum computer of moderate size is able to
compute the function $\alpha (g)$ for exponentially large values of $g$. 
One does not need to actually simulate the evolution of individual
points in the lattice, since $\alpha (g)$ is the smallest integer $t$
such that $L^{t}=I \; (\mbox{mod} \;g)\;$, with $I$ the
identity.  
Nine registers of $n_q$ qubits with $n_q \sim \log_2 g$
are needed, one holding the values of time,
the next four the matrix entries,
and the last ones being workspace.
The algorithm applies period-finding (which is the basis of 
Simon's \cite{simon}
or Shor's \cite{shor} algorithms)
to the four registers encoding the entries of the matrix.
First one prepares the initial state, which is
$N^{-1/2}\sum_{t=0}^{N-1}|t \rangle|1 \rangle|0 \rangle|0 \rangle|1 \rangle$ where $N=2^{n_q}$. 
It  is easily built from the ground state of the system by applying
$n_q$ Hadamard gates
and $2$ single-qubit flips.
Then one constructs the time evolution to end up with
 the entries of the matrix
$L^t$ modulo $g$ in the last four registers.
To this aim, first the entries $(a_i, b_i,c_i,d_i)$ 
of each $L^{2^i}$ for $i=1,...,n_q-1$
are precomputed classically.  This can be done sequentially, first squaring
$L$ to get $L^2$, then squaring $L^2$ and so on until $L^{2^{n_q-1}}$ is
obtained.
This requires of the order of $O(n_q^3)$ classical operations.
For each $t$, the binary decomposition of $t$ is
 $t=\Sigma_{i=0}^{n_q-1} \alpha_i 2^i$ with $\alpha_i=0,1$, so that
$L^t = \Pi_{i=0}^{n_q-1} (L^{2^i})^{\alpha_i}$.  So the exponentiation
of $L$ can be done by $n_q$ multiplication by $L^{2^i}$ conditioned 
on the value of the qubit $i$.
This matrix multiplication 
can be done by a sequence of number multiplication. For example, the
sequence of transformations for the first two entries is:
$N^{-1/2}\sum_{t=0}^{N-1}|t \rangle |a \rangle|b  \rangle|c \rangle|d \rangle$

$\rightarrow N^{-1/2}\sum_{t=0}^{N-1} |t \rangle |a \rangle|b \rangle|c \rangle|d \rangle|a \rangle|b \rangle$
(copy)

$\rightarrow N^{-1/2}\sum_{t=0}^{N-1} 
|t \rangle |aa_i \rangle|bd_i \rangle|c \rangle|d \rangle|ab_i \rangle|bc_i \rangle$

$\rightarrow 
N^{-1/2}\sum_{t=0}^{N-1} 
|t \rangle |aa_i+bc_i \rangle|ab_i+bd_i \rangle|c \rangle|d \rangle|ab_i \rangle|bc_i \rangle$.
This needs a controlled multiplier modulo $g$, which can be done following
the procedure in \cite{operation} using two extra registers as workspace.
To erase the unwanted registers one uses a standard trick:
one builds the sequence of gates corresponding to $F: |u \rangle|v \rangle|0 \rangle|0 \rangle \rightarrow
|u \rangle|v \rangle |d_ib_iu-c_ib_iv \rangle|a_ic_iv-b_ic_iu \rangle$.  
Applying $F^{-1}$ to the state above
sets the work registers to zero.  One then does again the same
operation for the next two entries of the matrix.  
This needs 
$O(n_q^2)$ operations, and should be done
for all qubits $i$ in the first register
to build $L^t$, so $O(n_q^3)$ gates are used in total.
One ends up with the state $N^{-1/2}\sum_{t=0}^{N-1}|t \rangle |A_t \rangle|B_t \rangle|C_t \rangle|D_t \rangle$
where ($A_t,B_t,C_t,D_t$) are entries of the matrix
$L^t$ modulo $g$.

Then as in Shor's algorithm  
one performs a Quantum Fourier Transform
(QFT) of the first register. This quantum analog of the 
classical Fourier transform needs $O(n_q^2)$ elementary gates
 to be performed. Since the last four registers describe a
periodic function of $t$, peaks in the Fourier
transform can be measured and yield the period $\alpha(g)$. 
Indeed, if $\alpha(g)$
divides $N$, then peaks are precisely at values of multiples of $N/\alpha(g)$,
while otherwise broader peaks will appear
but significant probability will be concentrated on 
the best rational approximants of multiples of $N/\alpha(g)$, and 
a continued fractions 
algorithm can yield 
the period in $O(n_q^3)$ classical operations \cite{josza,steane,nielsen}.

It is worth trying to compare this algorithm to compute $\alpha(g)$ with
others.  Classically, computing all iterates $L^k$ up to the period is
clearly exponential since $\alpha(g)$ is of the same order of magnitude as
$g$ ($\langle \alpha(g)/g \rangle \rightarrow 0$ slower than any 
inverse power of $g$
\cite{keating}).  Another method to 
compute
$\alpha(g)$ \cite{percival} is based on
the theory of ideals of quadratic fields.  For a prime $p$, $\alpha(p)=
\frac{p-(d/p)}{m}$,
where $()$ is the Kronecker symbol ($=\pm 1$), $d$ is the discriminant 
of the field containing the eigenvalues of $L$, and $m$ is an integer
 divisor of $p-(d/p)$ to be determined by trial and errors. For composite $g$,
$ \alpha(g)$ can be determined from the $\alpha(p)$, $p$ being the prime 
factors of $g$.  This method is also exponential classically, since it requires
the factorization of $g$ in prime factors.  On a quantum computer, one
may try to use 
Shor's algorithm, 
first factoring $g$, then factoring $p-(d/p)$ for all prime 
factors of $g$ and trying all possible integer divisors.  
Most integers $g$ have $\sim \log \log g$ prime factors 
and for most $p$ there are $\sim (\log p)^{\log 2}$ divisors of $p-(d/p)$
\cite{hardy}.
Thus this algorithm is also polynomial for most numbers.
Still, it is known \cite{hardy} that the number of divisors of an integer
$n$ can be quite large, of the order of $2^{\log n / \log \log n }$ 
in worst cases, making this method not polynomial for some $g$.
The algorithm here is
simpler, giving $\alpha(g)$ in one run, and always polynomial.

Another possibility to explore periodic orbits and Poincar\'e recurrences
which can be generalized to other systems beyond (\ref{catmap}),
is to start from an initial point
 and look for periodicities
of its iterates. In this case, it is appropriate to use a discretized 
map.  As discussed in \cite{ascat}, the best possible
discretization uses symplectic maps \cite{lattice}.
These maps can be exactly iterated and are themselves Hamiltonian
if the original system is.
In dimension $2$, they can be written as
the action of a unitary operator $L$ on the $N^2$ points $(x_i,y_j)$ with 
$x_i=i/N$, $i=0,...,N-1$ and $y_j=j/N$, $j=0,...,N-1$, with $N=2^{n_q}$.
 For example, for the cat map, a point $(x_i,y_j)$ of the discretized
phase space density is mapped to 
$(\bar{x_i},\bar{y_j})$ through the action of the $2\times 2$ matrix 
$L$ defined above. The algorithm requires
three registers, two of size $n_q$ specifying points in phase space
and one of size $p$ holding the values of time, plus additional 
qubits as workspace, with usually $p \approx n_q$.  
One starts from the initial state 
$2^{-{p/2}}\sum_{t=0}^{2^p-1} |t \rangle|x_0 \rangle |y_0 \rangle$,
easily built from the action of $p$ Hadamard gates
and $2n_q$ single-qubit flips.
Then one constructs a sequence of gates giving 
$2^{-{p/2}}\sum_{t=0}^{2^p-1}  |t \rangle|L^t(x_0) \rangle |L^t(y_0) \rangle$.
Then a QFT of the first register will yield peaks, 
from which the period of the point can be found.
If $L$ and each $L^{2^k}$ for $k=1,...,p-1$ can be computed
and implemented in $(O(B(n_q))$, where $B$ is a polynomial, this will be exponentially faster
than classically.
For the cat map, all can be done efficiently 
by a method similar to the one described
above, first precomputing classically the $L^{2^k}$,
then decomposing each $t$ in binary representation, 
and then using a controlled multiplier modulo $N$ to compute 
$(L^t(x_0),L^t(y_0))$.  Again, this uses $O(n_q^3)$
quantum gates. 

Other maps can be explored in the same way, provided their classical evolution
operator $L$ enables fast (polynomial)
classical computation of $(L^{2^k})$.  
In such cases, the algorithm above can reach exponential gain over the
classical computation.  If $L$ is efficiently implementable,
but exponentially large 
iterates cannot be efficiently computed (see examples below) the 
algorithm above will not work.
Nevertheless, in this case it is possible to obtain a gain for
the computation of periodic orbits and
recurrence properties of the whole system for a given time (as opposed
to a chosen initial point).
The basic idea is to compute sequentially the iterates of a 
distribution and then use Grover iterations to search for specific 
trajectories.
The dynamical system is discretized through a 
symplectic map $L$,
on a lattice of size $N \times N$ where $N= 2^{n_q}$, and it is assumed 
that $L$ can be implemented on these $N^2$ points in $O(B(n_q))$ operations
where $B$ is a polynomial. 
Then a subdomain $A$ is selected.
The simplest case is a square of size $P \times P$ with $P=2^p$ and $p< n_q$.
The initial state is 
$|\psi_0 \rangle = 2^{-{p}}\sum_{i=0}^{2^p-1} \sum_{j=0}^{2^p-1}|x_i \rangle |y_j \rangle$.
Then one realizes the transformation 
$|\psi_0 \rangle $
$\rightarrow$ $2^{-{p}}\sum_{i=0}^{2^p-1} \sum_{j=0}^{2^p-1}|L(x_i) \rangle |L(y_j) \rangle $
...$\rightarrow$ $2^{-{p}}\sum_{i=0}^{2^p-1} \sum_{j=0}^{2^p-1}|L^t(x_i) \rangle |L^t(y_j) \rangle $.
The value of the first $n_q-p$ qubits of each register
is enough to know if the trajectory is back to the
domain $A$ (since $A$ can be moved efficiently to a corner).
So after $t$ iterations this value is checked for all trajectories and a $\sigma_z$
controlled by the value of these $2n_q-2p$ qubits gives a minus sign to
the values of $|L^t(x_i)\rangle |L^t(y_j) \rangle$ which end a trajectory which returns to $A$.
This
can be done efficiently using $O((2n_q-2p)^2)$ elementary gates and no work
qubit \cite{nielsen}. 
Then one inverts all the gates but the last operation,
ending with 
$2^{-{p}}\sum_{i=0}^{2^p-1} \sum_{j=0}^{2^p-1}\epsilon |x_i \rangle |y_j \rangle$,
where $\epsilon=\pm 1$ depending on the fact that the trajectory went back to 
$A$ after these $t$ iterations. 
This whole procedure
is then used as an oracle for iterations of the Grover algorithm 
\cite{grover}.    
After $P/\sqrt{M}$ iterations, where $M$ is the number of solutions,
the amplitude of the wavefunction is concentrated on initial points
which return to $A$ after $t$ iterations,
and a measure of the registers will give one of them.
The whole algorithm necessitates $2n_q$ qubits plus the workspaces.
Only
$O(tP/\sqrt{M})$ operations are needed
to get the returns, as opposed to $O(tP^2/M)$ 
for the classical
simulation (up to logarithmic factors). 
In $O(tP/\sqrt{M})$ one initial point of a trajectory which returns to $A$
is obtained.
To get other trajectories which return, or for different return times,
one needs to restart the algorithm.  This process has obviously interest mostly for $t \ll P$.
The number of returning trajectories $M$ can
be evaluated through the quantum counting algorithm 
(phase estimation combined with the Grover iterations) in $O(tP)$ operations
\cite{counting}.  Since
generally the number $M$ is unknown, one can estimate it
by quantum counting
prior to the search, but it is not necessary \cite{counting}.

The algorithm can be modified to get a different result, 
which is the number and 
coordinates of the 
periodic orbits of the system. 
In this case one starts from all the $N \times N$ points of the lattice 
with $N=2^{n_q}$, with initial state
$|\psi_0 \rangle = 2^{-{n_q}}\sum_{i=0}^{2^{n_q}-1} 
\sum_{j=0}^{2^{n_q}-1}|x_i \rangle |y_j \rangle$.  
Then one performs:
$|\psi_0 \rangle$$\rightarrow$ $2^{-{n_q}}\sum_{i=0}^{2^{n_q}-1} 
\sum_{j=0}^{2^{n_q}-1}|x_i \rangle |y_j \rangle
|L(x_i) \rangle |L(y_j) \rangle $
...$\rightarrow$ $2^{-{n_q}}\sum_{i=0}^{2^{n_q}-1} \sum_{j=0}^{2^{n_q}-1}|x_i \rangle |y_j 
\rangle |L^t(x_i) \rangle |L^t(y_j) \rangle $.
After $t$ iterations the value of the iterate is compared to the initial value
(for example by adding the registers bitwise modulo $2$), 
and a minus sign is given (by using a $\sigma_z$
controlled by the value of these qubits \cite{nielsen})
if it is the same.  Then one invert all the gates but the last operation,
ending with $2^{-{n_q}}\sum_{i=0}^{2^{n_q}-1} 
\sum_{j=0}^{2^{n_q}-1}\epsilon |x_i \rangle |y_j \rangle $,
where $\epsilon=\pm 1$ depending on the fact that the trajectory is 
periodic of period $t$ or not. 
This whole procedure
is then used as an oracle for iterations of the Grover algorithm \cite{grover}
 and all periodic orbits can be found
by iterating the process.  After $O(tN/\sqrt{M})$ operations,
the amplitude
of the wavefunction is concentrated on periodic points of period $t$
($M$ is the number of such points).
In contrast, the classical search needs $O(tN^2/M)$ operations.
As above $M$ 
can be evaluated through quantum counting in 
$O(tN)$ operations.

A much studied class of classical twist maps corresponds to
the form  $\bar{n} = n - K V'(\theta)(\mbox{mod}\; 2\pi L) ;\;
\bar{\theta} = \theta + \bar{n} (\mbox{mod} \; 2\pi)$ 
where $(n,\theta)$ is the pair of conjugated momentum (action) and angle 
variables, and the bars denote the resulting variables after one iteration 
of the
map.  The discretized map on a $N \times N$ lattice is simply 
$\bar{Y} = Y + [N K V'(2 \pi X/N)/(2 \pi) ] (\mbox{mod} N) ;\;
\bar{X} = X + \bar{Y} (\mbox{mod} N)$ where $[... ]$ is the integer part 
and $X, Y$ are integers.
The case $V(\theta)=\cos\theta$ corresponds to the Chirikov standard map
which has been a cornerstone model in the study of chaos \cite{lieberman}.  
The discretized map on a $2^{n_q}\times 2^{n_q}$ lattice
can be implemented in $O(n_q^3)$ gates 
on a quantum computer \cite{BG1} and therefore
a quadratic gain can be achieved for return times and periodic orbits
(although other quantities may be obtained with exponential gain \cite{ascat}).
A simpler example is the 
sawtooth map which corresponds to $V(\theta)= -\theta^2/2$.
The discretized mapping is $\bar{Y} = Y + [N K (2 \pi X/N-\pi)/(2 \pi) ] 
(\mbox{mod} N);\;
\bar{X} = X + \bar{Y} (\mbox{mod} N)$. 
 This model has been intensively studied in the field
of classical chaos \cite{saw}.
  Depending on the values of $K$, the system can be stable
or chaotic, and can display complex structures in phase space with 
chaotic and integrable parts. Interesting phenomena such
as anomalous diffusion, self-similar structures, etc... are present.
 For integer $K$, exponentially large iterates can be computed
efficiently and the gain is exponential.  For non integer $K$, the
iterates cannot be computed efficiently but the map itself can be implemented
in $O(n_q)$ gates on a $2^{n_q}\times 2^{n_q}$ lattice, and
a quadratic gain can be achieved for return times and periodic orbits.
  For example, the case $K=\pm 1/2$ 
for return times 
requires only three registers (two for $X$ and $Y$, one as workspace)
and the algorithm can be realized with 
as little as 8 qubits and 40 gates per Grover iteration
(domain $4\times 4$ in a $8 \times 8$ lattice).

A particular case arises for systems with symmetries.  For example,
in the standard and sawtooth maps, the classical evolution operator $L$ is
the product of two involutions $L=I_1I_2$ with $I_1^2=I_2^2=1$.
Such involutions have lines of fixed points, and it is easy to 
show that trajectories crossing twice one of these lines are periodic.
They correspond to periodic orbits which are invariant 
through one of the symmetries.  These orbits can be found by iterating 
only points from one of these lines, i.e. $N$ points instead of $N^2$.
  The latter algorithm finds these 
orbits in $\sqrt{N}$ operations, keeping the quadratic speed-up.

The periodic orbits found by the various algorithms above are exact 
periodic orbits of the discretized systems, which is a Hamiltonian system in
its own right.  As concerns the original continuous system,
for hyperbolic systems 
the shadowing theorem \cite{shadowing} ensures
that an exact trajectory will remain close to the dynamics of each
discretized point for arbitrary times.  This exact trajectory will return
close to its starting point after the time computed by the algorithm, 
meaning that it is a Poincar\'e recurrence time of the system.
Moreover, such approximately periodic trajectory can be used as a
starting point for a Newton method converging to a true periodic orbit.

In conclusion, it has been shown that on a quantum computer one can obtain
Poincar\'e recurrence times and periodic orbits of certain classical dynamical
systems exponentially faster that on a classical computer.  For a larger class
of systems, a quadratic gain can be achieved.

I thank D. Shepelyansky for many very useful discussions, 
and S. Bettelli and M. Terraneo for critical reading of the manuscript.
This work was supported in part 
by the EC  RTN contract HPRN-CT-2000-0156 and by the EC
project EDIQIP of the IST-FET programme.


\vskip -0.5cm
\begin{thebibliography}{99}
\bibitem{feynman} R.~P.~Feynman, Found. Phys. {\bf 16}, 507 (1986)
\bibitem{josza} A.~Eckert and R.~Josza, Rev. Mod. Phys. {\bf 68}, 733 (1996).
\bibitem{steane} A.~Steane, Rep. Progr. Phys. {\bf 61}, 117 (1998).
\bibitem{nielsen} M.~A.~Nielsen and I.~L.~Chuang, {\em Quantum computation
                  and quantum information}, Cambridge Univ. Press, 2000.
\bibitem{shor} P.~W.~Shor, in Proc. 35th Annu. Symp. Foundations of
               Computer Science (ed. Goldwasser, S. ), 124 
               (IEEE Computer Society, Los Alamitos, CA, 1994).
\bibitem{grover} L.~K.~Grover, Phys. Rev. Lett. {\bf 79}, 325 (1997).
\bibitem{lloyd} S.~Lloyd, Science {\bf 273}, 1073 (1996).
\bibitem{BG1} B.~Georgeot and D.~L.~ Shepelyansky, Phys. Rev. Lett.
                 {\bf 86}, 2890 (2001).
\bibitem{saw1} G.~Benenti, G.~Casati, S.~Montangero and D.~L.~Shepelyansky, 
                 Phys. Rev. Lett. {\bf 87}, 227901 (2001).
\bibitem{lidar} D.~A.~Lidar and O.~Biham, Phys. Rev. E {\bf 56}, 3661 (1997).
\bibitem{ascat} B.~Georgeot and D.~L.~ Shepelyansky, Phys. Rev. Lett. 
              {\bf 86}, 5393 (2001); Phys. Rev. Lett. 
              {\bf 88}, 219802 (2002).
\bibitem{Poincare} H.~Poincar\'e, {\em Les m\'ethodes nouvelles de la
                  m\'ecanique c\'eleste}, Gauthier-Villars, Paris (1892).
\bibitem{recurrences} C.~Karney, Physica {\bf 8D}, 360 (1983);
                     B.~V.~Chirikov and D.~L.~Shepelyansky,
                     {\em ibid.} {\bf 13D}, 395 (1984);
                      J.~Meiss and E.~Ott, Phys. Rev. Lett. {\bf 55},
                      2741 (1985); G.~Zaslavsky and M.~Tippett,
                      {\em ibid.} {\bf 67}, 3251 (1991);
                      B.~V.~Chirikov and D.~L.~Shepelyansky, 
                      {\em ibid.} {\bf 82}, 528 (1999).
\bibitem{cvitanovic} D.~Auerbach, P.~Cvitanovic, J.-P.~Eckmann, 
                    G.~H.~Gunaratne and I.~Procaccia, Phys. Rev. Lett. 
                    {\bf 58}, 2387 (1987); P.~Cvitanovic,
                      Phys. Rev. Lett. {\bf 61}, 2729 (1988). 
\bibitem{GTF} M.~C.~Gutzwiller, {\em Chaos in classical and quantum
                     mechanics}, Springer, N. Y. (1990).
\bibitem{arnold} V.~I.~Arnold and A.~Avez, {\em Ergodic Problems of Classical 
               Mechanics}, Benjamin, N. Y. (1968).
\bibitem{lieberman} A.~Lichtenberg and M.~Lieberman, {\em Regular and Chaotic 
                  Dynamics}, Springer, N.Y. (1992).
\bibitem{hannay} J.~H.~Hannay and M.~V.~Berry, Physica {\bf 1D}, 267 (1980).
\bibitem{percival} F.~Vivaldi, Proc. Roy. Soc. London {\bf A 413}, 97 (1987);
                    I.~C.~Percival and F.~Vivaldi, Physica {\bf 25D}, 
                   105 (1987).
\bibitem{keating} J.~P.~Keating, Nonlinearity {\bf 4}, 277 (1991).
\bibitem{rudnick} P.~Kurlberg and Z.~Rudnick, Commun. Math. phys. {\bf 222},
                  1, 201 (2001). 
\bibitem{simon} D.~Simon, in Proc. 35th Annu. Symp. Foundations of
               Computer Science (ed. Goldwasser, S. ), 116 
               (IEEE Computer Society, Los Alamitos, CA, 1994).
\bibitem{operation} V.~Vedral, A.~Barenco and A.~Ekert,  Phys. Rev. A {\bf 54},
                      147 (1996); D.~Beckman, A.~N.~Chari, S.~Devabhaktuni 
                      and J.~Preskill,
                       Phys. Rev. A {\bf 54}, 1034 (1996);  
                     C.~Miquel, J.~P.~Paz and R.~Perazzo, 
                      Phys. Rev. A {\bf 54}, 2605 (1996).
\bibitem{hardy} G.~H.~Hardy and E.~M.~Wright, {\em An introduction to the 
                 theory of numbers}, Oxford University Press, Oxford 
                 ($5^{th}$ Ed., 1979).
\bibitem{lattice} F.~Rannou, Astron. Astrophys. {\bf 31}, 289 (1974);
               D.~J.~D.~Earn and S.~Tremaine, Physica D {\bf 56}, 1 (1992).
\bibitem{counting}M.~Boyer, G.~Brassard, P.~H{\o}yer and A.~Tapp, 
                 Fortschr. Phys. {\bf 46}, 493 (1998).
\bibitem{saw} I.~Dana, N.~W.~Murray and I.~C.~Percival, Phys. Rev. Lett.
                       {\bf 62}, 233 (1989); Q.~Chen, I.~Dana, J.~D.~Meiss,
                       N.~W.~Murray and I.~C.~Percival, Physica {\bf 46 D},
                       217 (1990).
\bibitem{shadowing} D.~V.~Anosov, Proc. Steklov. Inst. Math. 
                    {\bf 90},1 (1967); R.~Bowen J. Diff. Eq. {\bf 18}, 333 
                     (1975); T.~Sauer, C.~Grebogi and J.~A.~Yorke, 
                    Phys. Rev. Lett. {\bf 79}, 59 (1997).
\end{thebibliography}
\end{document}